\documentclass[aps,prb,twocolumn,superscriptaddress]{revtex4-1}

\usepackage{bm}        
\usepackage{amsmath,amssymb,amsthm}   
\usepackage{mathtools}
\usepackage{braket}
\usepackage{tabularx}

\usepackage{hyperref}

\usepackage{grffile}
\usepackage{graphicx}
\usepackage{float}
\begin{document}
\title{Terahertz Spin-Orbital Excitations in the paramagnetic state of multiferroic Sr$_2$FeSi$_2$O$_7$}
\author{Thuc T. Mai}
\affiliation{Center for Emergent Materials, Department of Physics, The Ohio State University. Columbus, OH 43210, USA}
\author{C. Svoboda}
\affiliation{Center for Emergent Materials, Department of Physics, The Ohio State University. Columbus, OH 43210, USA}
\author{M.T. Warren}
\affiliation{Center for Emergent Materials, Department of Physics, The Ohio State University. Columbus, OH 43210, USA}
\author{T.-H. Jang}
\affiliation{Department of Physics, Pohang University of Science and Technology, Pohang 37673, Korea}
\affiliation{Laboratory for Pohang Emergent Materials and Max Plank POSTECH Center for Complex Phase Materials, Pohang University of Science and Technology, Pohang 37673, Korea}
\author{J. Brangham}
\affiliation{Center for Emergent Materials, Department of Physics, The Ohio State University. Columbus, OH 43210, USA}
\author{Y.H. Jeong}
\affiliation{Department of Physics, Pohang University of Science and Technology, Pohang 37673, Korea}
\author{S-W. Cheong}
\affiliation{Laboratory for Pohang Emergent Materials and Max Plank POSTECH Center for Complex Phase Materials, Pohang University of Science and Technology, Pohang 37673, Korea}
\affiliation{Rutgers Center for Emergent Materials and Department of Physics and Astronomy, Rutgers University, Piscataway, New
Jersey 08854, USA}
\author{R. Vald\'es Aguilar}
\email{valdesaguilar.1@osu.edu}
\affiliation{Center for Emergent Materials, Department of Physics, The Ohio State University. Columbus, OH 43210, USA}
\date{\today}

\begin{abstract} 
We studied the novel multiferroic material Sr$_2$FeSi$_2$O$_7$, and found 3 absorption modes above the magnetic ordering transition temperature using time-domain terahertz spectroscopy. These absorption modes can be explained as the optical transitions between the spin-orbit coupling and crystal field split 3d$^6$ Fe$^{2+}$ ground state term in this material. Consideration of the compressed tetrahedral environment of the Fe$^{2+}$ site is crucial to understand the excitations. We point out, however, discrepancies between the single-site atomic picture and the experimental results.
\end{abstract}

\maketitle
Mutiferroics belong to the class of materials that host more than one type of ferroic order, i.e. (anti)ferroelectricity, (anti)ferromagnetism, ferroelasticity, etc. In the presence of multiple types of ordering, for example magnetization \textbf{M} and electric polarization \textbf{P}, exotic properties can emerge. It has been demonstrated that \textbf{P} can be controlled by an external magnetic field, and \textbf{M} can be controlled by an external electric field in TbMnO$_3$ and TbMn$_2$O$_5$ \cite{Kimura2003,Hur2004,Cheong-Mostovoy}. Exotic optical properties such as directional dichroism and magneto-chiral dichroism have been observed in the multiferroic Ba$_2$CoGe$_2$O$_7$ (BCGO)\cite{Bordacs2012,Kezsmarki2011} as well. In BCGO, \textbf{M} and \textbf{P} are coupled below the ordering transition temperature T$_{Neel}$ $\sim$ 6 K. This coupling results in an absorption mode that is both electrically and magnetically active \cite{Kezsmarki2011,Aguilar2009,Pimenov2006}. An interesting feature of these phenomena is that the electrically-active magnon (or electromagnon) mode survives above the Neel temperature, which has so far not received a full explanation. The electromagnon is a signature of the multiferroic phase, and is one of the most exciting discoveries in the area of quantum magnetism in the last ten years.

It was recognized early in the multiferroic renaissance that the Dzyalonshinskii-Moriya (DM) interaction is crucial for the coupling between ferroelectricity and spiral magnetic order \cite{Mostovoy-spiral,Sergienko-DMI,KNB2005}. The origin of the DM interaction is the spin-orbit coupling (SOC) \cite{Dzya,Moriya1,Moriya2}, and thus SOC is at the center of the phenomena of magnetically induced ferroelectricity. Although SOC can also generate dynamical effects \cite{KBN2007}, the dynamical response of electromagnons in the multiferoic families of RMnO$_3$ and RMn$_2$O$_5$ (where R is a rare-earth ion) has been explained by the symmetric Heisenberg exchange-striction \cite{Sushkov-JPCM,Aguilar2009} without the need for SOC. One of the exceptions is in fact BCGO, where SOC has to be explicitly taken into account to explain the electromagnons in that system \cite{Romhanyi-2011,Soda-2014,Murakawa-2010,Penc-2012,Miyahara-2011,Romhanyi-2012}. Surprisingly, however, these theories have not included the effect of tetragonal distortion of the CoO$_4$ tetrahedra in the explanation of the static and dynamical properties of BCGO. This omission is even more glaring as highlighted by the fact that the tetragonal distortion of the tetrahedra is of 13\% compression with respect to a perfect tetrahedron. For comparison, we find that the compression in SFSO is $\sim$ 17\%. Just recently the effect of this distortion on the electronic properties of BCGO and similar materials has been studied using first-principles calculations \cite{Picozzi-BCGO,Baraone-JahnTeller}.

In this paper, we report the terahertz (THz) study above 7.5 K of Sr$_2$FeSi$_2$O$_7$ (SFSO), a material isostructural to BCGO (space group \#113, P$\bar{4}$2$_1$m). The single crystal SFSO samples show multiple absorption modes above the magnetic order transition temperature, T$_{Neel}\sim$ 5 K \cite{Cheongprivate}. These absorption modes can be understood as transitions between the spin-orbit and crystal field split ground state levels of the Fe$^{2+}$ ion on a compressed tetrahedral environment. We find it crucial to consider this tetragonal distortion in order to explain the details of these magnetic excitations. We will discuss the shortcomings of this model and point to potential avenues for a better understanding.

\begin{figure*}[t]
\includegraphics[width=1\textwidth]{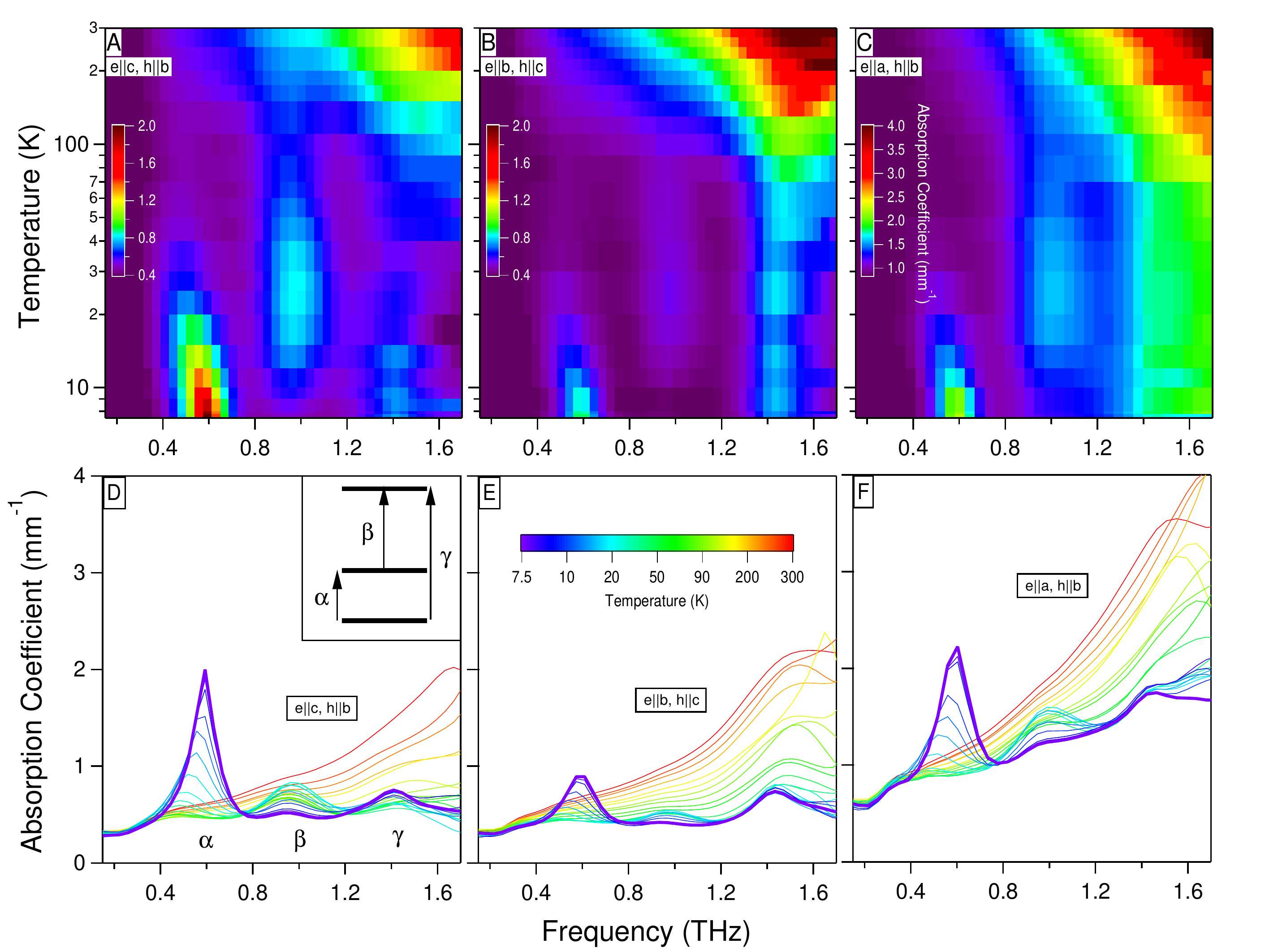}
\caption{\textbf{Spin-Orbital THz absorption of Sr$_2$FeSi$_2$O$_7$}.
False color maps of the absorption from 7.5 K to 300 K and below 1.7 THz for the (A) e$_\omega\parallel$c, h$_\omega\parallel$b, (B) e$_\omega\parallel$b, h$_\omega\parallel$c , and (C) e$_\omega\parallel$a, h$_\omega\parallel$b  configurations. The color corresponds to the absorption value as indicated in the scale bar in each individual panel. (D)-(F) Selected absorption traces for the same polarization configurations as in (A)-(C), respectively. The color scale bar in (E) applies to the three (D) through (F) panels. Inset in (D) shows the three level structure that is apparent in the temperature dependence of the absorption. Transitions are labeled as $\alpha$, $\beta$, and $\gamma$ as indicated. A different absorption mode appears at higher T, and dominates the other three.}
\label{fig:SFSOAbs}
\end{figure*}

We used a home-built time-domain terahertz spectrometer (TDTS) with photoconductive antennas as source and detector of THz radiation. This technique has recently risen to the forefront of the study of novel excitations in quantum magnets \cite{Morris-columbite,Laurita2015,pan2014low}. It has the advantage of being of high energy resolution for Brillouin zone-center excitations, and it does not need as large single crystals as other techniques. TDTS also has the advantage of a phase sensitive measurement, which means that one can obtain the complex optical constants of the material.~The samples were mounted inside a closed-cycle cryostat with optical access windows that is capable of cooling to 7 K. By comparing the frequency components of a THz pulse that has passed through the sample to one reference without a sample, we can extract the transmission coefficient, $t(\omega)$, as a function of frequency. The absorption coefficient can then be extracted from the transmission data as -$\log{t(\omega)}/d$, where $d$ is the thickness of the crystal. In this experiment, we fitted each absorption peak with a Lorentzian lineshape, characterized by the absorption peak frequency, its full width at half maximum, and by its spectral weight. 

We studied two single crystals of SFSO: one is $a$-plane cut $\sim$1150 $\mu m$ thick; and the second one, $\sim$460 $\mu m$ thick, is $c$-plane oriented. In order to clarify the nature of the absorption modes with their selection rules, we used a wire grid polarizer to linearly polarize the THz pulse along the different crystalline axes. To avoid birefringence in the $a$-plane sample, we measured the sample with the THz electric field, e$_\omega$ polaried parallel to the b axis, and with e$_\omega\parallel$c. As expected, the $c$-plane sample did not show any birefringence due to its tetragonal crystal structure.  The single crystals were grown using a floating zone method in a reducing atmosphere with feed rods prepared through a solid-state reaction.

\begin{figure}[t]
\includegraphics[width=1\columnwidth]{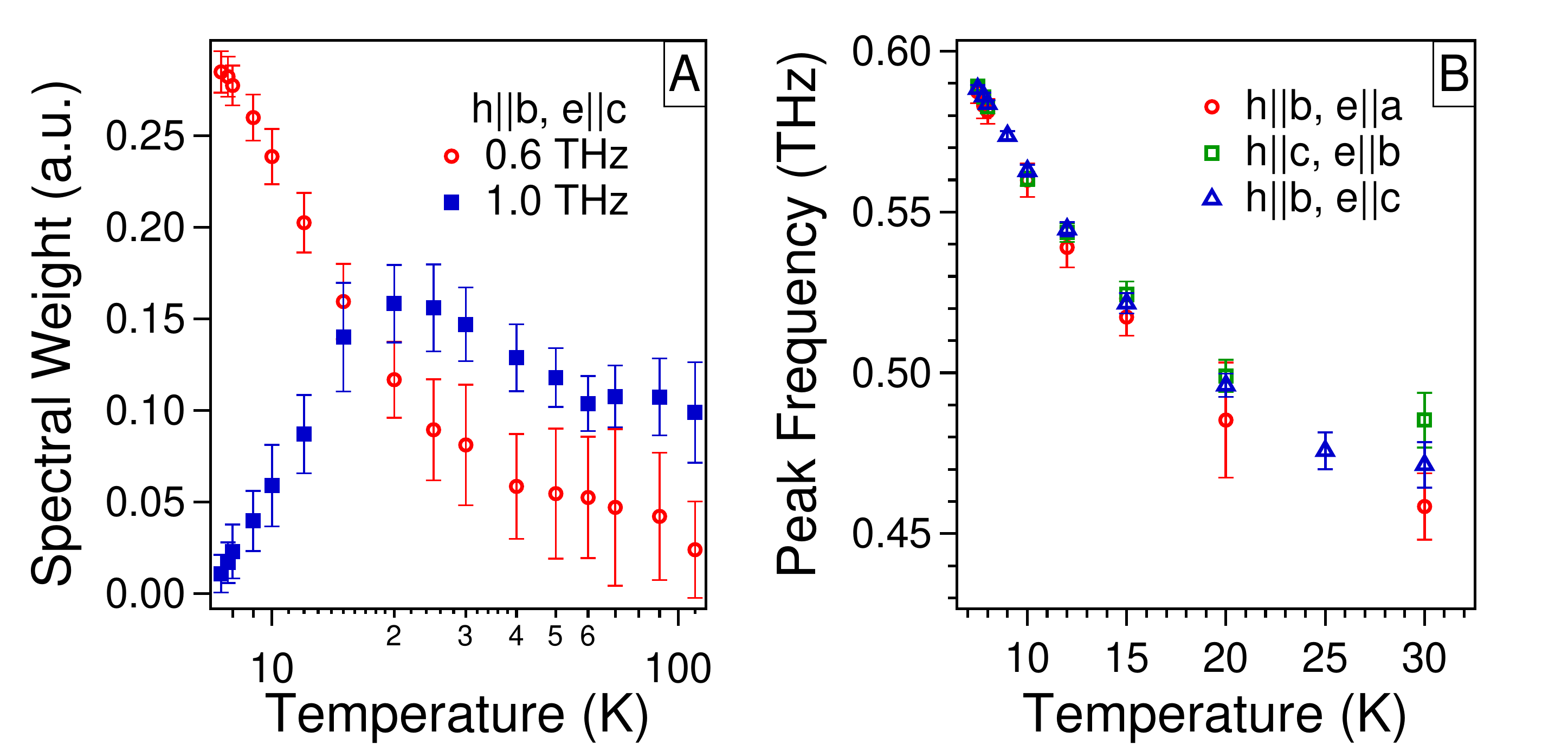}
\caption{\textbf{Temperature dependence of transition parameters}. (A) Spectral weight of $\alpha$ and $\beta$ transitions versus temperature in the polarization configuration $e_\omega\parallel$c, $h_\omega\parallel$b. The $\beta$ transition clearly loses all its spectral weight towards zero temperature, a clear indication of a transition between two excited states. (B) Temperature dependence of the frequency of the $\alpha$ transition for the three polarization configurations explored in this work. Within error bars, the frequencies are identical, a signature of the same transition appearing in all polarization configurations.}
\label{fig:fig2}
\end{figure}

\begin{figure}[t]
\includegraphics[width=1\columnwidth]{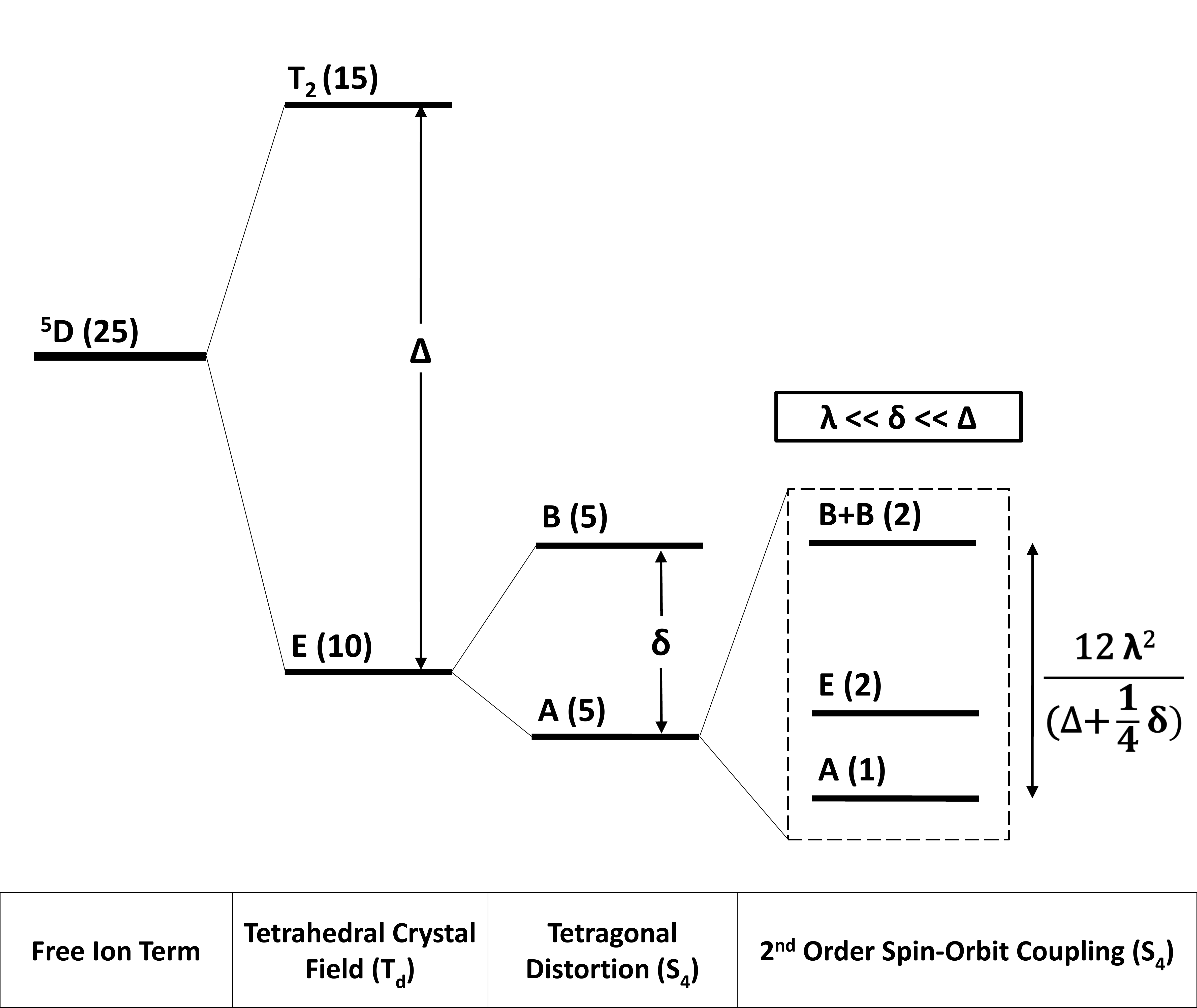}
\caption{\textbf{Ground state splitting by crystal field and spin-orbit coupling}. Splitting of the Fe$^{2+}$ term ($^5$D) in the S$_4$ crystal field environment and due to spin-orbit coupling up to second order in perturbation theory. The numbers in parenthesis indicate the degeneracy of the states. The full splitting is only shown for ground state. States are labeled by the irreducible representations of the point group corresponding to the distortion. Energy separations are not to scale. The full term splitting can be found in \citet{Low1960}.}
\label{fig:EnergyDiagram}
\end{figure}

We measured the absorption spectra of Sr$_2$FeSi$_2$O$_7$ from 7.5 to 300 K, and we identify three measurable absorption modes with frequencies of $\sim$ 0.6, 1.0, and 1.4 THz, as shown in Figure:\ref{fig:SFSOAbs}(D), labeled as $\alpha$, $\beta$, and $\gamma$, respectively. At 7.5K, the $\alpha$ mode is the strongest among these with absorption coefficient around a few tens of cm$^{-1}$, comparable to BCGO's electromagnon mode \cite{Kezsmarki2011}. Meanwhile, the $\beta$ mode is virtually absent at 7.5 K but gradually increases in strength and peaks at 20 K. The $\gamma$ mode at 1.4 THz is the weakest of the three. It is clear from the behavior of the spectral weight of the $\beta$ mode (see figures \ref{fig:SFSOAbs}(D)-(F), and figure \ref{fig:fig2}(A)) that it corresponds to a transition between two excited states, as its strength first increases and then decreases with temperature, a clear indication of population of the excited states with temperature. The $\alpha$ and $\gamma$ absorptions, on the other hand, are clearly from a ground state to two different excited states. This phenomenology can be captured in a simple three-level system depicted in the inset of figure \ref{fig:SFSOAbs}(D) and in figure \ref{fig:EnergyDiagram}. The details will be discussed below.

Around 100 K, another absorption mode starts to move down into our frequency range, completely dominating the spectra at room temperature. We believe this high frequency-high temperature mode is consistent with a polar phonon. Interestingly, the peak positions of the $\alpha$ and $\beta$ modes are red shifted with increasing temperature, cf. fig. \ref{fig:fig2}, a behavior typical of an order parameter-type phase transition. However, there is no known phase transition above $\sim$ 5 K in this material. Therefore, our simple phenomenological three-level system will need to be expanded in order to explain these frequency shifts.

All three modes exists in all orientations of the THz electric field e$_{\omega}$ with respect to the crystal axes, see figures \ref{fig:SFSOAbs}(D)-(F). This suggests that these modes are active under both the electric and magnetic dipole selection rule. This is similar to the behavior of the low energy excitations found in BCGO \cite{Kezsmarki2011}. We find this behavior in the paramagnetic state of SFSO as opposed to the magnetically ordered one in BCGO. It is also the case, however, that the 1 THz electromagnon in BCGO survives to temperatures higher than the Ne\'el temperature, and has been suggested to be a \textit{spin-stretching} mode \cite{Penc-2012}. In SFSO, the main features of the polarization selection rules of the three modes are: 1) when the magnetic field of the THz is in the $a-b$ plane,  h$_\omega\parallel b$, the $\alpha$ mode has the same intensity regardless of the direction of the THz electric field  e$_\omega$, 2) this also applies to the $\gamma$ mode. 3) When h$_\omega\parallel c$ or e$_\omega\parallel a-b$ plane, all three modes are much weaker than in the other two orientations. This behavior suggests that the modes are mainly of magnetic dipole character in the $a-b$ plane, h$_\omega\parallel b$, and are only weakly electric dipole on the same plane, e$_\omega\parallel b$. This is summarized in table \ref{table1}.

\begin{table}[t]
\caption{\textbf{Selection rules for the excitation of the observed absorption modes}. The number of checkmarks, \checkmark, indicates the strength of the absorption for the given THz polarization configuration. Here e$_\omega$ and h$_\omega$ are the electric and magnetic fields of the THz pulse, respectively. a, b and c are the crystallographic axes.}
\label{table1}
\setlength{\extrarowheight}{4pt}
\noindent\begin{tabularx}{0.87\linewidth}{| c | c | c | c |}
\hline
 & h$_\omega\parallel$ b \& e$_\omega\parallel$ c & h$_\omega\parallel$ b \& e$_\omega\parallel$ a
& h$_\omega\parallel$ c \& e$_\omega\parallel$ b \tabularnewline
\hline
$\alpha$ &\checkmark\checkmark\checkmark &\checkmark\checkmark\checkmark &\checkmark \tabularnewline
$\beta$ &\checkmark\checkmark &\checkmark\checkmark &\checkmark \tabularnewline
$\gamma$ &\checkmark &\checkmark &\checkmark \tabularnewline
\hline
\end{tabularx}
\end{table}

We can begin understanding the nature of these excitations by utilizing a single site picture of the Fe$^{2+}$ ion in the crystal field environment of a compressed tetrahedron. \citet{Low1960} showed how the energy levels of the Fe$^{2+}$ ion are split due to cubic crystal fields, including tetrahedral symmetries. These predictions were somewhat confirmed by \citet{Slack1967} in the THz range where Fe$^{2+}$ occupies a tetrahedral site in a ZnS matrix. In this case, however, Fe$^{2+}$ ions are very diluted and do not interact with each other. Recent interest has been given to the excitations of a regular lattice of tetrahedrally coordinated Fe$^{2+}$ in the material FeSc$_2$S$_4$ \cite{Mittelstadt2015,Laurita2015}. It is thought that this material does not magnetically order, however next-nearest neighbor exchange has been used to theoretically explain the experimental results \cite{Ish2015}. The effect of this exchange interaction between Fe$^{2+}$ sites is to strongly renormalize the energies of the 5-fold split ground state term, and it also gives a dispersion in momentum space to the otherwise dispersionless single-site excitation.

Fe$^{2+}$ has a 3d$^6$ (L=2, S=2) electronic configuration in free space ($^5$D term), but in Sr$_2$FeSi$_2$O$_7$ it occupies the 2a Wyckoff position in the P$\bar{4}$2$_1$m space group that has S$_4$ ($\bar 4$) site symmetry. This site symmetry corresponds to 4 O$^{2-}$ ions located at the vertices of a tetrahedron compressed along the [001] crystallographic direction. We performed single crystal X-ray diffraction on our samples and obtained a compression of approximately 17\%. We therefore model the electronic structure, following \citet{Low1960}, assuming an energy hierarchy of $\Delta\gg\delta\gg\lambda$, where $\Delta$ is the E--T$_2$ tetrahedral crystal field splitting, $\delta$ is the tetragonal compression splitting A--B, and $\lambda$ is the spin-orbit interaction energy. In this limit, in the high-spin configuration and following Hund's rules, the ground state is a spin-orbital singlet $\ket{E_0}$ (eqn. \ref{eq1}), and the first two excited states are doublets up to second order in $\lambda$, $\ket{E_1}$ (eqn. \ref{eq2}) and $\ket{E_2}$ (eqn. \ref{eq3}). Their wavefunctions to first order in $\lambda$ are (the basis for this expansion are the states $\ket{L_z,S_z}$, where both $L_z$ and $S_z$ go from -2 to +2):
\begin{equation}\label{eq1}
\ket{E_0} = \ket{0,0} + x \left( \sqrt{6}\ket{+1,-1} + \sqrt{6}\ket{-1,+1} \right) + \mathcal{O}(\lambda^2)
\end{equation}
\begin{equation}\label{eq2}
\ket{E_1^\pm} = \ket{0,\pm 1} + x \left( \sqrt{6} \ket{\pm 1,0} + 2 \ket{\mp 1,\pm2} \right) + \mathcal{O}(\lambda^2)
\end{equation}
\begin{equation}\label{eq3}
\ket{E_2^\pm} = \ket{0,\pm 2} + x \left( 2\ket{\pm 1, \pm 1} \right) + \mathcal{O}(\lambda^2)
\end{equation}
\noindent where $x=\frac{\sqrt{6}\lambda}{8(\Delta+\delta/4)}$, and $x\ll1$. Note that the largest contribution to each of the states is derived from the $L_z$ = 0 submanifold, and the $S_z$ spin contribution changes by $\pm$1 in each of the 2 excited states. This is derived from the fact that the lowest energy d-orbital, the $z^2$ orbital \cite{note1}, is doubly occupied as given by the crystal field of the compressed tetrahedron \cite{Picozzi-BCGO,Baraone-JahnTeller}. We note that for a perfect tetrahedron, the ground state manifold is split into five equally separated energy levels by the crystal field and SOC \cite{Low1960}.

The second order spin-orbit Hamiltonian takes the form $DS_z^2$, which is the typical form for single ion anisotropy energy, where $D=\frac{\lambda^2}{16(\Delta+\delta/4)}$, we take as the zero of the energy the state $\ket{E_0}$. In this particular case, since $D\geq0$, this is an easy-plane anisotropy. Figure \ref{fig:EnergyDiagram} schematically shows the splitting of the ground state term by the crystal field and spin-orbit coupling \cite{Low1960}. The states are now labeled by the irreducible representations of the point groups belonging to each level of distortion, where SOC does not break any symmetry. We note that, whereas $\ket{E_1^\pm}$ is a doublet of E symmetry, $\ket{E_2^\pm}$ are two distinct states of B symmetry; they are accidentally degenerate only up to second order in $\lambda$. We can obtain a value for $\lambda$ using $\Delta\approx$ 0.8 eV and $\delta\approx$ 0.1 eV \cite{khomskii2014}, and the predicted energy separation between $\ket{E_0}$ and $\ket{E_2}$, $\frac{12\lambda^2}{\Delta+\delta/4}$, and obtain an upper limit of $\lambda\approx$ 20 meV.

Thus, the single-ion picture already contains a three-level structure that reproduces the basics of the experimental observations. Transitions between the states $\ket{E_0}$ and $\ket{E_1}$, and between $\ket{E_1}$ and $\ket{E_2}$ are magnetic dipole allowed as they are connected by an operator $S_\pm = S_x\pm iS_y$ that changes the $S_z$ value by one. Therefore, a THz magnetic field polarized in the $a-b$ plane is able to make transitions between these states. The transition between $\ket{E_0}$ and $\ket{E_2}$, in this approximation, is only electric quadrupole since $\Delta S_z=\pm 2$ (terms of $\mathcal{O}(\lambda^3)$ and higher make the transition between $\ket{E_0}$ and $\ket{E_2}$ magnetic dipole as well). However, we note that because of the lack of inversion symmetry of the Fe$^{2+}$ site, parity is not a good quantum number for the wavefunctions, admixtures of the $^5$D ground term with higher energy terms of different parity (i.e. P and F terms) \cite{Low1960} are allowed. This mixing will make all the transitions between the three lowest states weakly electric dipole as well, as we find experimentally.

As we noted above, a feature of the data that cannot be explained by this single-ion picture is the fact that the lowest transition frequency has a very strong temperature dependence, whereas the second and third transition frequencies barely change, see figure \ref{fig:fig2}(B). Therefore, the single-ion picture would need to be expanded to include the effects of interactions between the spin and orbital angular momenta of the Fe$^{2+}$ ions at different sites. In addition, we highlight again that all of this phenomenology is occurring above the magnetically ordered transition temperature, and thus we expect that models such as those of \citet{Ish2015} would be required to explain all of our experimental results.  Below the ordering temperature, we expect to see antiferromagnetism of a similar type as BCGO. Magnetic ordering would lower the symmetry of the system even further, potentially causing splittings and further shifts in the absorption spectrum.

In summary, we have observed three spin-orbital transitions in the range between 0.2 THz and 1.7 THz in Sr$_2$FeSi$_2$O$_7$ using time-domain THz spectroscopy. These modes can be qualitatively explained by Fe$^{2+}$ in a compressed tetrahedron crystal field environment where spin-orbit coupling splits the low energy manifold into a three-level structure with a singlet ground state and two doublet excited states. We estimate a spin-orbit coupling constant for Fe$^{2+}$ of $\lambda\approx$ 20 meV. We find it crucial to consider the effects of the compression of the tetrahedron, as without it, the low energy manifold would consist of five equally spaced energy levels \cite{Low1960}. The effect of this compression should be taken into account when explaining the THz excitations observed in BCGO as well. We also find that, although the single-ion picture can qualitatively explain many of our results, modifications will be needed to explain the strong shift with temperature of the transition frequency between the first two states. Measuring the THz absorption below T$_{Neel}$ and under an applied magnetic field will shed more light onto the nature of the low energy excitations in Sr$_2$FeSi$_2$O$_7$, measurements which are now underway.

We acknowledge the assistance of R.D. Dawson, E.V. Jasper, and K. Meng in carrying out the THz measurements, and to Dr. Judith Gallucci for help with the X-ray diffraction measurement. Work at OSU was supported in part by The Ohio State University, and by the Center for Emergent Materials, an NSF MRSEC under grant DMR-1420451. The work at Rutgers University was supported by the DOE under Grant No. DOE: DE-FG02-07ER46382. Work at Postech was supported by the Max Planck POSTECH/KOREA Research Initiative Program [Grant No. 2011-0031558] through NRF of Korea funded by MSIP, and YHJ acknowledges support from NRF Korea through grant 2015R1D1A1A02062239.

\bibliography{SFSOBib}

\begin{thebibliography}{33}
\bibitem [{not()}]{note1}%
  \BibitemOpen
  \href@noop {} {}\bibinfo {note} {To avoid confusion, the single electron
  orbitals are labeled with lower-case letters, i.e. d-orbitals, whereas the
  multi-electron ground term is labeled with the upper-case letter
  D.}\BibitemShut {Stop}%
\end{thebibliography}

\end{document}